# Medical Applications

*C. Biscari and L. Falbo*
ALBA-CELLS, Barcelona, Spain
CNAO Foundation, Pavia, Italy

**Abstract**
The use of accelerators for medical applications has evolved from initial experimentation to turn-key devices commonly operating in hospitals. New applications are continuously being developed around the world, and the hadrontherapy facilities of the newest generation are placed at the frontier between industrial production and advanced R&D. An introduction to the different medical application accelerators is followed by a description of the hadrontherapy facilities, with special emphasis on CNAO, and the report closes with a brief outlook on the future of this field.

## 1 Introduction

The development of particle accelerators has been driven by fundamental research, but their applications in society have been growing since the very beginning of their history. About half of the particle accelerators in the world are now used for medical purposes. This amounts to several thousand instruments, ranging from small cyclotrons, to linacs for radiotherapy, up to the large infrastructures created for hadrontherapy.

Among the most common uses we can mention is the production of radioisotopes, mainly for diagnostic purposes and oncological therapies.

Cancer can be considered to be one of the most widespread diseases around the world. It is the second major cause of death (after cardiovascular diseases) in developed countries, where the risk of contracting a form of cancer before the age of 75 years is estimated to be around 25–30% [1]. Cancer is the hysterical and irregular growth and propagation of a cluster of cells. The technique of radiotherapy is based on the principle of using ionizing particles to damage the DNA of the cancer cells in order to kill them or block their ability to regenerate.

About seven million people are known to die each year because of this disease, and a similar number of patients are successfully cured, which means they experience a symptom-free survival period exceeding five years.

Soon after their discovery in 1895, X-rays were used for medical purposes in the treatment of diseased tissues. From these early completely empirical tests, radiotherapy has evolved to become an important tool in medicine and one of most exploited techniques in the fight against cancer.

Classical radiotherapy employs photons and electrons, produced by electron linacs, whereas hadrontherapy uses protons and light ions. The first hadrontherapy treatments were performed in clinically adapted particle physics research centres; today there are dedicated facilities designed and built as hadrontherapy clinical centres. The realization of machines for hadrontherapy is more challenging than for standard radiotherapy; whereas many hospitals have a device for classical radiotherapy, hadrontherapy needs a dedicated complex, hosting a cyclotron or a synchrotron, and the related treatment rooms.

It is estimated that about 40 million patients have been treated with photon therapy, which has a long tradition; hadrontherapy has been developed more recently, and the number of treated patients with protons and carbon ions is about 100 000 at the time of writing (2013), with a rate of increase that approaches 10% per year.

The medical accelerator field is involved in robust R&D programs. The development of increasingly compact devices, based on obtaining high accelerating gradients and simplifying beam delivery to the patients, is in progress in several institutions around the world, involving several communities, one of the most active of which is the high-energy-physics community.

In this report, we briefly introduce activities associated with radioisotope production. Section 3 will be dedicated to classical radiotherapy. The core of the report will discuss hadrontherapy, with special emphasis on synchrotron-based facilities, of which CNAO will be taken as an example. Finally, we mention the R&D underway for future facilities.

## 2  Radioisotope utilization

Nuclear medicine utilizes radioactive substances for diagnosis and therapeutic purposes.

### 2.1  Diagnostics

Imaging using radioisotopes can be divided into two main techniques: Single-Photon Emission Computed Tomography (SPECT) and Positron Emission Tomography (PET) [2].

Short-lived isotopes are used for SPECT: they emit gamma rays from within the body, where they can be introduced by injection, by inhalation, or orally. The gamma rays are detected by gamma cameras, and the images are used to reconstruct the emitting organs and indicate abnormal conditions.

A radioisotope used for diagnosis must emit gamma rays of sufficient energy to escape from the body, and it must have a half-life short enough for it to decay away soon after imaging is completed.

Technetium-99m is the radioisotope most widely used in medicine. It is employed in about 80% of all nuclear medicine procedures—some 30 million per year, of which 6–7 million are in Europe, 15 million are in North America, 6–8 million are in Asia/Pacific (particularly Japan), and 0.5 million are in other regions. It is an isotope of the artificially produced element technetium, and it has almost ideal characteristics for a nuclear medicine scan [3, 4].

Positron emission tomography is a precise technique which uses positron-emitting radionuclides, which are injected and accumulated in the target tissue. As they decay, they emit a positron, which combines with a nearby electron, resulting in the simultaneous emission of two identifiable gamma rays in opposite directions. They are then detected by a PET camera, which gives a very precise indication of their origin. The most important clinical role of PET is in oncology, which uses fluorine-18 as the tracer, since it has proven to be the most accurate non-invasive method of detecting and evaluating most cancers. It is also often used in cardiac and brain imaging.

Radioisotopes are produced mainly through nuclear reactions in reactors or from charged particle bombardment in accelerators.

In accelerators, the typical charged particle reactions utilize mainly protons, but also deuterons and helium nuclei ($^3$He$^{2+}$ and alpha particles). Most radionuclides are produced in cyclotrons from the bombardment of enriched molybdenum. The inventor of the cyclotron, Ernest Lawrence, used deuterons to bombard a carbon target in 1934 at Berkeley and induced a reaction that resulted in the formation of a radioisotope with a half-life of 10 min.

Recently, small low-energy linacs and tandem cascade accelerators have been used in radioisotope production. The development of such devices is driven by the recent shortage of radioisotopes from nuclear reactors, due to the shutting down of several of the long-lived facilities. The advantages of accelerators with respect to reactors are the fact that they do not produce radioactive waste, and have few radionuclide impurities. Furthermore, the target and the product are different chemical elements, and are more easily separable.

## 2.2 Therapy

Gamma beams from a radioactive cobalt-60 source are also used for teletherapy, i.e. external irradiation, although this technique is almost everywhere being replaced by the much more versatile linear accelerators as high-energy X-ray sources, and these will be described later.

Gamma knife radiosurgery focuses gamma radiation from cobalt-60 sources on a precise area of the brain with a cancerous tumour. Worldwide, over 30 000 patients are treated annually, generally as outpatients.

In internal radionuclide therapy a small radiation source, usually a gamma or beta emitter, is administered or planted in the target area. Short-range radiotherapy is known as brachytherapy, and this is becoming the primary means of treatment. Iodine-131, a very successful cancer treatment, is commonly used to treat thyroid cancer as well as non-malignant thyroid disorders. Iridium-192 implants are used especially in the head and breast. This brachytherapy (short-range) procedure presents less overall radiation to the body, is more localized to the target tumour, and is cost effective.

## 3  General aspects of radiotherapy

Radiotherapy is used to control a tumour locally and, in some situations, in the surrounding diffusion paths. A dose of ionizing particles is delivered to the tumour, the target for the radiation beam.

Photons, electrons, protons, and ions can be and are used for this purpose, but neutrons are rarely used. The dose profile delivered to tissues of each type of beam is represented by the Bragg curve (see Fig. 1).

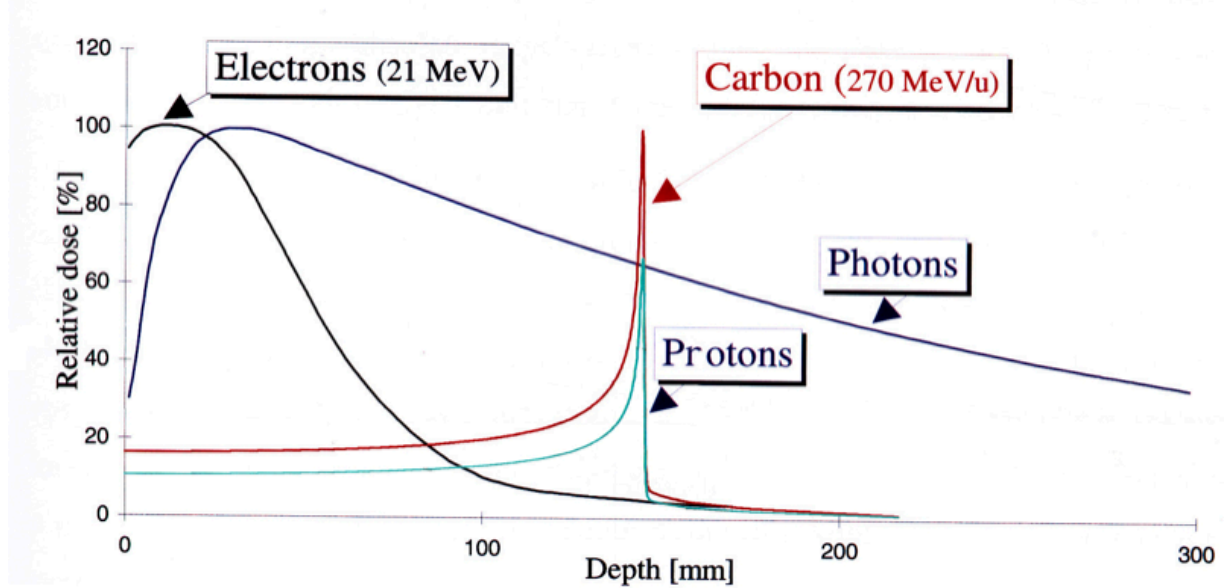

**Fig. 1:** Qualitative depth dependence of the deposited dose for each radiation type, with the narrow Bragg peak at the end corresponding to hadrons [5].

Electron beams deposit most of their energy in the surface region, as in tissue they have a limited range (which depends on the beam energy), beyond which only a very low-intensity tail generated by bremsstrahlung photons is present. Due to these characteristics, electron beams are suitable for the treatment of superficial or semi-deep tumours, and are used in about 10% of all conventional treatments.

X-rays and gamma rays are characterized by an exponential attenuation, following a maximum reached after a few centimetres, corresponding to the maximum range of the secondary electrons produced by the primary photons in the most superficial layers of the irradiated tissue. As a consequence of this build-up effect, for high-energy photon beams the skin dose is relatively low.

Neutrons have a similar profile, and are sometimes used in treatments.

For proton and light ion beams, the dose is lower at the surface and presents a peak at the depth where the particles stop (known as the *Bragg peak*). This characteristic allows the radiation dose to be effectively concentrated so it can treat deep-seated tumours efficiently. The radiation field can be shaped not only transversally, but also longitudinally using several Bragg peaks at different penetration depths that create the so-called SOPB (Spread-Out Bragg Peak, see Fig. 2). The hadrons mostly exploited are protons and carbon ions.

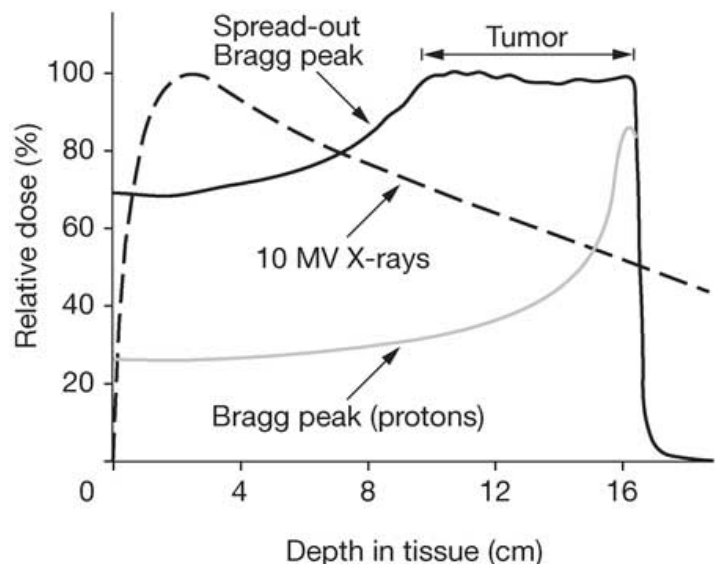

**Fig. 2:** Spread-out Bragg peak [6]

The radiation beams have other figures of merit that account for the different biological and clinical effects of the interaction of the dose with the tissues.

The first is the Linear Energy Transfer (LET, symbol $L_\infty$) [7] or *stopping power*, which is the density of energy deposition along the track of the particles forming the radiation field.

The LET expresses, at the microscopic level, the modalities of energy transfer from the radiation to the tissues, and the unit used in practice is the kiloelectronvolt per micrometre (keV/$\mu$m). The index $\infty$ on $L$ indicates that there is no limitation on the amount of energy released in any single collision of the charged particle with an atom or molecule of the traversed medium. For X-rays the relevant charged particles are the electrons put in motion by the beam, mainly through Compton effects, for which $L_\infty$ is in the typical range 0.2–2 keV/$\mu$m. Given these low values, electrons and X-rays are referred to as *sparsely ionizing radiation*. On the other hand, one millimetre before stopping in matter the LET of protons is about 10 keV/$\mu$m. Carbon ions, fully stripped of their electrons, have a LET which is on average about 25 times larger than for protons of the same range. For cobalt gamma rays the maximum LET is about 10 keV/$\mu$m, for protons it is approximately 100 keV/$\mu$m, and for heavier ions it may reach 1000 keV/$\mu$m, presenting a high value in the Bragg peak region and a low one at the beam entrance.

Another parameter is the Relative Biological Effectiveness (RBE) [8], defined as the ratio of the absorbed dose of a reference radiation (generally the 2 MeV gamma photons emitted by cobalt-60) and that of the test radiation required to produce the same biological effect, quantified by the cell survival fraction on the same tissue.

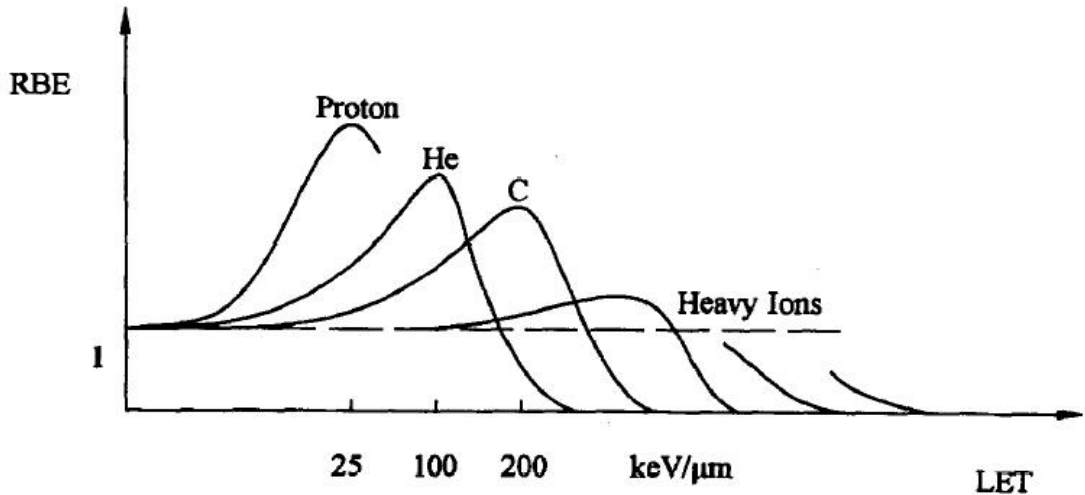

**Fig. 3:** RBE as a function of LET

For all therapeutic X-ray beams, RBE = 1, while for hadrons the value of RBE for a given cell type and a defined effect (endpoint) depends mainly on the LET value, although it also varies with the ion species chosen, as shown in Fig. 3.

Finally, the Oxygen Enhancement Ratio (OER) [9] is the ratio of the dose necessary to produce a biological effect in the absence of oxygen and the dose necessary to produce the same effect in the presence of oxygen. The photon OER is about 3; it decreases when the LET is greater than 100 keV/$\mu$m, and approaches unity at 300 keV/$\mu$m.

Another aspect to be considered is multiple scattering: for higher-mass ions, the scattering is less relevant, leading to an improvement in the lateral and longitudinal dose distribution but to an increase in the nuclear fragmentation, which produces a tailing of the Bragg peak.

Theoretical studies, taking into account all these aspects, indicate that ions with $Z > 6$ should not be a good clinical choice. When, during the 1980s hadrontherapy had a revival in Europe and Japan, carbon ions were indicated as the best medical choice, and were often the only solution for radio-resistant tumours. Other species in the range $1 < Z \leq 6$ could be as or more interesting than carbon ions [10], and clinical experiments at existing hadrontherapy facilities could reveal interesting results.

## 4 Linacs for X-ray radiotherapy

X-rays used in radiotherapy are produced by the interaction of an accelerated electron beam with a heavy-metal target. High-energy X-rays are shaped conforming to the tumour shape, either by blocks placed at the exit of the machine or by multileaf collimators.

The patient treatment couch is movable in all directions, and the beam can be rotated around the patient using a gantry.

Compact electron linacs, accelerating electrons to energies of the order of 10 MeV and focusing them onto the target, are now industrially produced and extensively used in hospitals around the world.

Major advances have been obtained with Intensity-Modulated Radiation Therapy (IMRT). It provides a computer-controlled precise radiation dose conforming to the three-dimensional shape of the tumour by modulating the intensity of the radiation beam in multiple small volumes, using hundreds of tiny radiation collimators.

Treatment is planned by using 3-D Computed Tomography (CT) or Magnetic Resonance Imaging (MRI) of the patient together with computerized calculations to determine the dose intensity pattern that will best conform to the tumour shape. Typically, combinations of multiple intensity-modulated fields coming from different beam directions produce a customized radiation dose that maximizes tumour dose while also minimizing the dose to adjacent normal tissues.

Currently, IMRT is extensively used to treat cancer of the prostate, head and neck, and central nervous system [11]. IMRT has also been used in limited situations to treat breast, thyroid, and lung, as well as in gastrointestinal organs, gynaecologic malignancies, and certain types of sarcomas. IMRT may also be beneficial for treating paediatric malignancies.

The development of combined therapies includes the possibility of using both X-rays and electron beams produced by the same linac [12].

Radiation can also be administered during surgery: Intraoperative Radiation Therapy (IORT) [13] is an intensive radiation used to treat cancers that are difficult to remove during surgery and where there is a concern that microscopic amounts of cancer may remain. IORT allows direct radiation to the target area while sparing normal surrounding tissue, and also allows higher effective doses of radiation to be used compared to conventional radiation therapy, which in several cases cannot afford very high doses due to the vicinity of sensitive organs. It is often combined with conventional radiation therapy usually administered before surgery.

## 5  Hadrontherapy

Hadrontherapy is based on the use of hadrons for irradiating the diseased tissues. The hadrons commonly used are protons and heavy ions with $Z < 6$. The advantage of hadrons over electrons and photons is easily explained by the Bragg curve, as shown in Fig. 1. The hadron Bragg curve is characterized by a narrow peak that occurs at a distance from the beam entrance: this provides good dose localization, with a low dose at both the entrance and the exit of the tumour target.

### 5.1  Hadrontherapy facility design criteria

#### 5.1.1  *Choice of accelerator*

The considerations reported in Section 3 are of fundamental importance when defining the design of a hadrontherapy centre. The main points that influence the characteristics of such a facility are the ion species to be accelerated and the technique used to shape the radiation field. Three different accelerator types are possible: linear accelerators, cyclotrons, and synchrotrons.

The penetration depth ranges between 30 mm and 300 mm. In the case of protons and carbon ions, this corresponds to the energy ranges 60 MeV–220 MeV and 120 MeV/u–425 MeV/u, respectively. In principle, these energy ranges can be obtained with all three accelerators. However, linacs are not very practical or feasible for high energies, and therefore we will consider only cyclotrons and synchrotrons, which are the main layouts in the hadrontherapy facilities. On the other hand, synchrotrons can perform easily the acceleration of both proton and carbon ions. Considering that the limitation is the magnetic rigidity, a synchrotron for carbon ions can accelerate all species with $1 \leq Z \leq 6$; also, oxygen can be accelerated with such a layout, but only up to a penetration range of 190 mm. Even if they are more flexible than cyclotrons, synchrotrons are technologically more complicated and therefore more costly; for example, the synchrotron needs an injection energy of several MeV/u, which requires an injector linac. The cyclotron appears to be more compact, especially in the case of a superconducting one. In the case of proton beam acceleration, a cyclotron has a diameter of about 4–5 m, whereas a synchrotron reaches 7–11 m if accelerating protons, and up to about 20 m for carbon ions.

The maximum energy of carbon ions makes the realization of a dedicated cyclotron very challenging: to date, cyclotrons for 400 MeV/u carbon ions have not been realized, but a centre has been recently proposed by IBA [14] consisting of a carbon cyclotron and a proton cyclotron. The advantage of more compact accelerators is partially reduced by the overall size of the facility, which is occupied mainly by the beam lines and the treatment rooms with the gantries and the technical infrastructures. The current from the cyclotron is d.c., while in a synchrotron it is pulsed because of the need to ramp the magnets from the injection value to the extraction value first, and then to a maximum value that allows avoiding non-repeatability problems when changing energy due to the magnetic hysteresis. As a consequence, currents from cyclotrons are generally much higher than those from synchrotrons: in the case of protons, cyclotrons can deliver about 300 nA instead of a few nanoamps from synchrotrons.

### 5.1.2  *Beam delivery*

There are essentially two techniques used to shape beam distribution on the tumour target: passive and active beam delivery.

*Passive* delivery essentially consists in placing before the patient several absorbers able to change beam characteristics. The main elements of the system, which are sketched in Fig. 4, are as follows:

− a scatterer to enlarge the beam;

− a variable degrader and a ridge filter to increase energy spread creating a SOBP;

− a first collimator to select the central part of the beam;

− the so-called 'bolus', a device with a 'hole' that has the shape of the distal surface of the tumour;

− a final multileaf collimator that gives the beam the required transverse size.

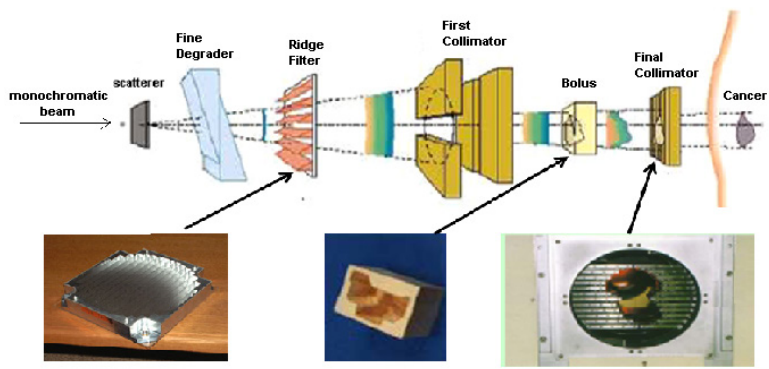

**Fig. 4:** Schematic of passive scanning

Some variants of this scheme are the use of a rotating wheel range modulator as a variable degrader and the wobbling method. The rotating wheel allows the thickness of the material that the beam passes through to be changed; by rotating the wheel, beams with different energies are obtained, resulting in a SOBP. The wobbling method is based on the use of scanning magnets that cause the beam to move on a circle at high frequency before the scatterer, resulting in a flat beam to be adjusted transversally and longitudinally.

There are some obvious disadvantages of the passive method. First, the bolus and the multileaf collimator are strongly dependent on the tumour, and also they are specific for each patient. Second, as shown in Fig. 5, the bolus takes into account only the distal surface, causing the proximal parts of the tumour to be very highly irradiated. Third, the presence of lots of materials between the beam and the patient causes nuclear fragmentation that leads to dose tails after the Bragg peak. In particular, in the case of heavy ions, passive scanning has other drawbacks. Heavy ions cause nuclear fragmentation with the target; furthermore, since they scatter less than protons, thicker scatterers are needed to obtain a large treatment field: thicker scatterers imply higher energy and beam losses requiring higher energies and currents from the accelerators.

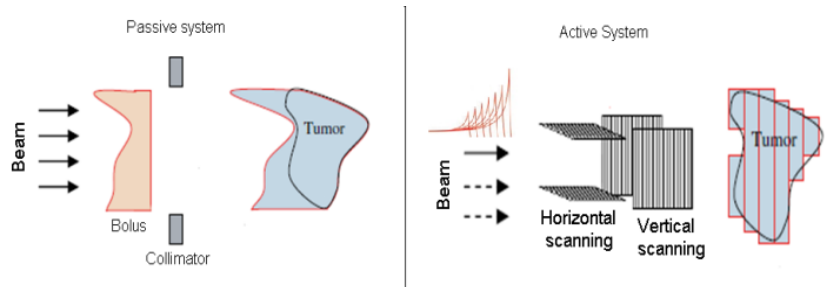

**Fig. 5:** Dose uniformity in the case of passive and active systems

*Active* scanning was first used in Japan in 1980 [15], and was then optimized and regularly used for treatments at PSI [16], GSI [17], HIT, and CNAO [18].

In the active scanning method, two magnets are used to move the beam in the two orthogonal transverse directions. The tumour is virtually divided into slices in the longitudinal direction, and each slice is thought of as composed of small volumes called voxels (or spots). Each slice is irradiated by fixing the beam energy and irradiating each voxel, which changes the currents of the scanning magnets. Furthermore, for each voxel in a slice the dose given during the irradiation of the previous slices can be taken into account.

Therefore with active scanning there is no specific hardware for each patient, and the whole irradiated target is very close in shape to the tumour target in both the transverse and the longitudinal planes. The drawback of such a beam delivery system is its greater complexity in operation due to the management of the scanning magnets and beam position, and also an increased sensitivity of the system to current ripples and changes. The tumour must be known with the same precision that characterizes the active scanning to obtain the required precision in the dose shaping. Problems occur when the tumour moves because of the patient's breathing and heartbeat. In this case passive scanning appears to be the easier solution; considering the superiority of the active method, several studies are in progress worldwide to develop methods that permit the use of active scanning with moving tumours: repainting, gating, and beam tracking [19]. Repainting [20] consists of treating the target multiple (about 10) times with a reduced dose; in this way, the amplitude, period, and initial phase of the organ motion change randomly treatment by treatment, and the irradiation uncertainty is statistically reduced. Gating [21] is a technique that is used with passive scanning. It is based on the irradiation of the tumour only during a precise percentage (about 30%) of the organ motion; in this way cyclotron treatment increases proportionally, whereas in a synchrotron this disadvantage is mitigated by the cycle times needed to fill the ring. Finally, beam tracking [22] is an adjustment of the parameters of the treatment plan in real time using a 4D organ monitoring signal. A purely active scanning method, i.e. without absorbers, is possible only with a synchrotron because of the need for variable extraction energy. Indeed, the energy from a cyclotron is fixed, and active scanning is possible only after having changed the beam energy, as in the passive methods, with a wedge degrader (resulting in a maximum energy variation rate of about 15 MeV/s).

## 5.2 Hadrontherapy worldwide

The idea of hadrontherapy appeared in 1946 in a paper written by Robert Wilson [23] that proposed the medical use of protons produced by the new high-energy accelerators. His idea was first realized when 30 patients were treated with protons at the Lawrence Berkeley Laboratory (LBL) in 1954. In the following years, other treatments have been performed in other research centres worldwide, such as in Uppsala in 1957 and at Harvard in 1963. Proton therapy followed in new facilities that became operative in Russia (Dubna in 1967, Moscow in 1969 and St. Petersburg in 1975), in Japan (Chiba in 1979 and Tsukuba in 1983), and in Switzerland at the PSI centre in 1985.

The world's first hospital-based dedicated proton facility started treatments in 1990, 20 years after the feasibility study at Loma Linda. The LLUMC (Loma Linda University Medical Center) synchrotron has a diameter of 6 m with a 2 MeV injector placed on top of the ring. A beam of $2\cdot10^{10}$ particles per spill is extracted in the range 70–250 MeV with a half-integer resonant extraction scheme. The facility is equipped with a fixed beam room with two beam lines (for eye and head-and-neck treatments), three rotating gantries, and a research room with three beam lines. To date, over 15 000 patients have been treated.

At the present time, 45 hadrontherapy facilities are in operation all around the world. Figure 6 shows the locations of the facilities. Most centres are proton facilities using cyclotron technology with passive beam delivery system, with fewer hadrontherapy synchrotrons.

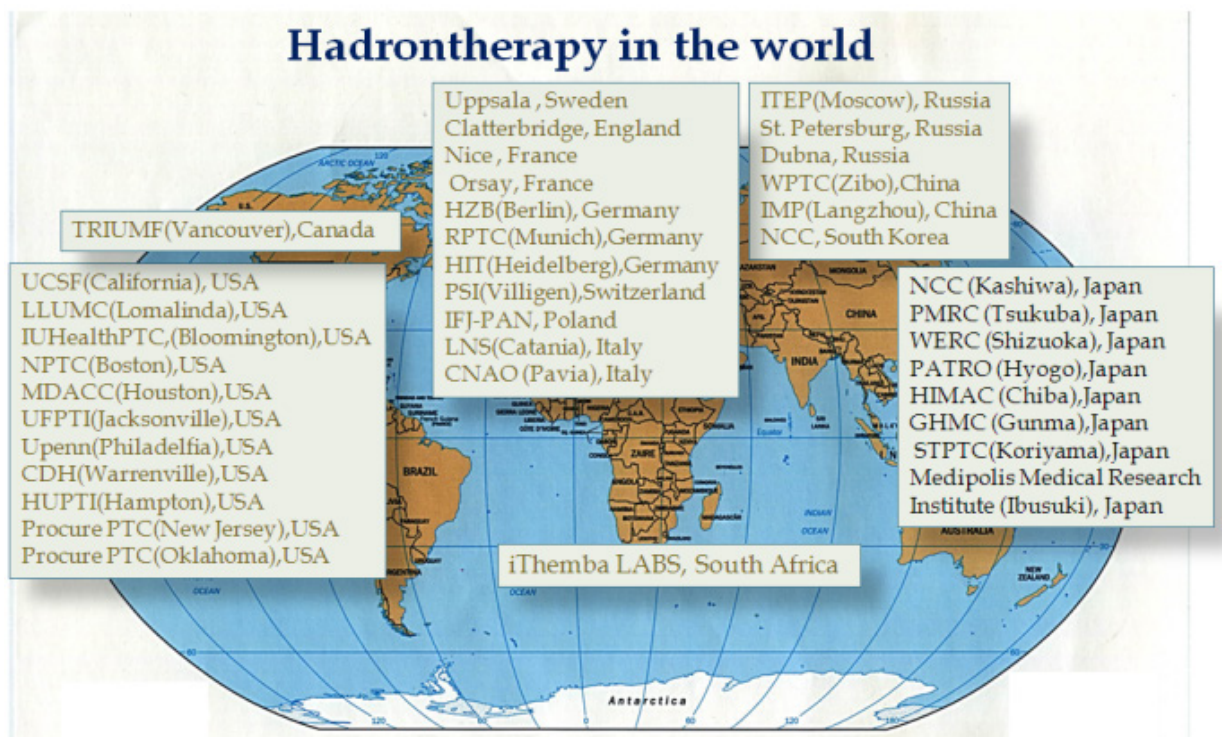

**Fig. 6:** Locations of the operative hadrontherapy facilities all around the world

Other facilities are under construction or will begin treatments over the next few years. Presently, carbon ions are produced in Asia at HIMAC, GHC, PATRO, and IMP, and in Europe at HIT and CNAO. Among these facilities, PATRO [24], HIT [25], and CNAO [26] produce both protons and carbon ions, and HIMAC, GHC, and IMP are dedicated facilities using only carbon ions. Particular mention should be made of the HIMAC centre [27] (see Fig. 7): in operation since 1994, it produces the most important clinical results with carbon ions (more than 7000 patients treated). It is equipped with two synchrotrons on the upper and lower floors that are much larger than others (42 m diameter instead of the standard 20–22 m) because they were designed to deliver 800 MeV/u Si ions for clinical experimentation.

Private companies, such as IBA, Hitachi, and Mitsubishi, are now able to commercialize proton beam therapy centres based on cyclotrons with largely standard components, including the treatment rooms and related gantries. A few synchrotron based centres also providing carbon ions have been produced by Mitsubishi and Sumimoto, but the commercialization is still not standard.

Up to December 2013, the total number of patients treated with hadrons in the world was almost 100 000, of which about 10% have been treated with carbon ions [28].

The number of centres and treated patients in the last few decades has grown exponentially, as shown in Fig. 8.

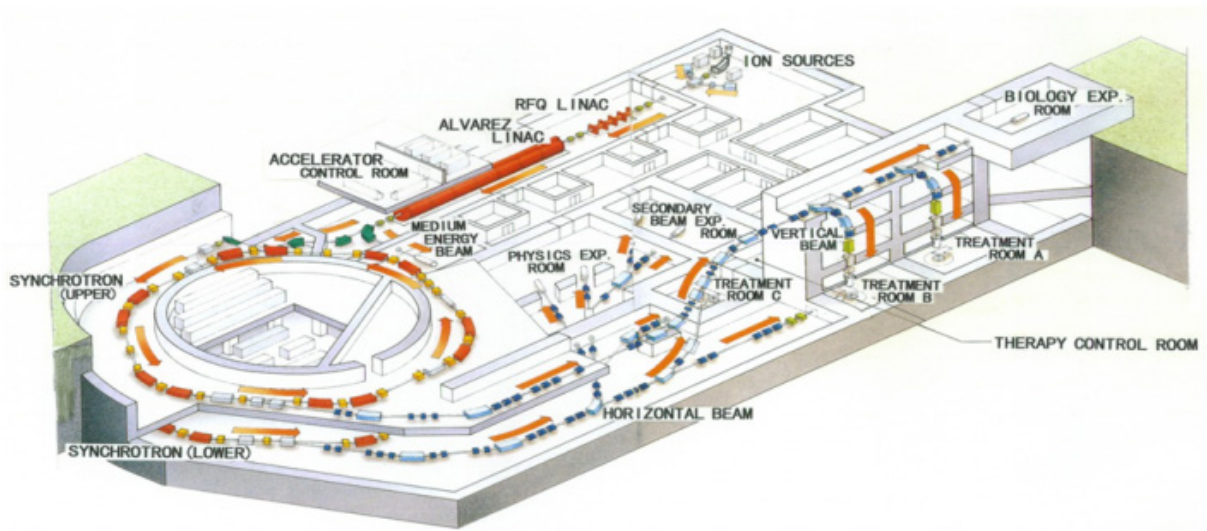

**Fig. 7:** HIMAC layout

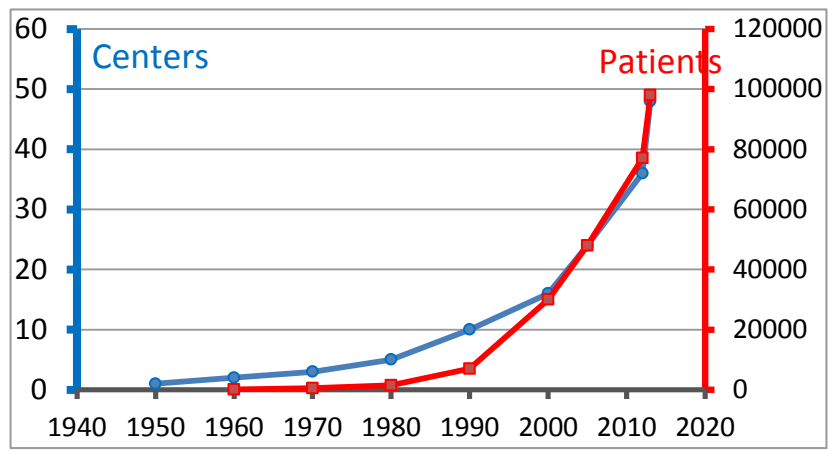

**Fig. 8:** Growth of hadrontherapy centres and treated patients in the last 60 years

### 5.3 Layout of a typical synchrotron

This section will show in detail some aspects of a synchrotron facility that is able to deliver proton and carbon ions. To provide some typical orders of magnitude, the CNAO facility [26], shown in Fig. 9, will be used as an example.

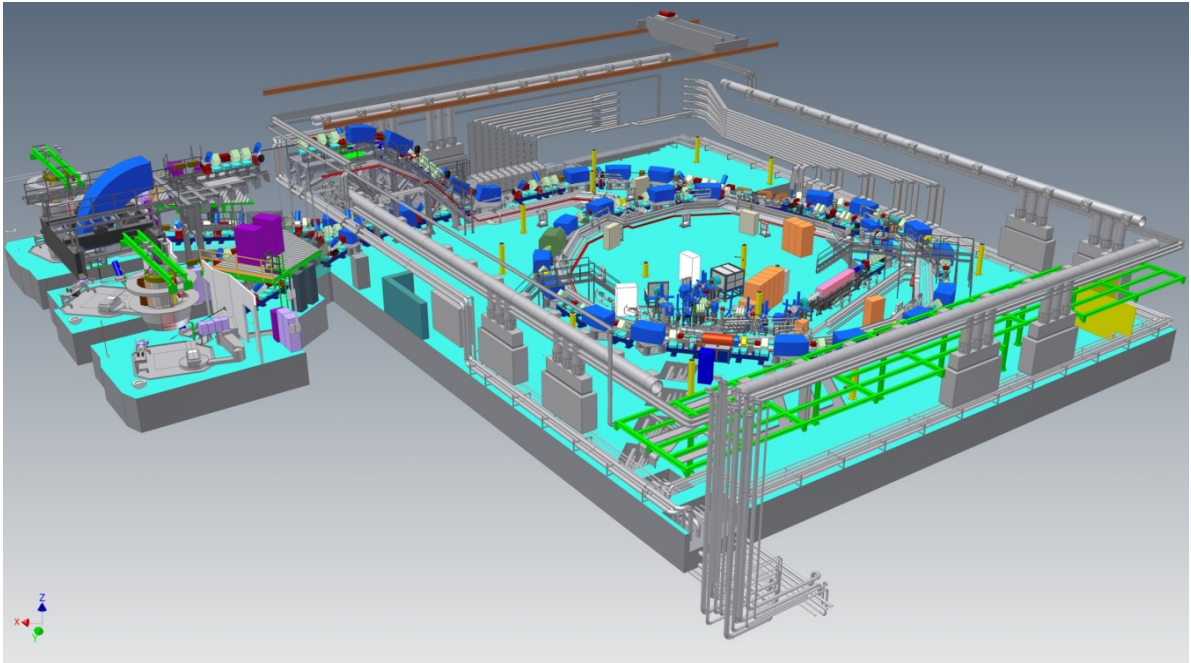

**Fig. 9:** CNAO, 3D rendering

The layout of such a facility comprises a low-energy injector, a synchrotron ring, and several beam lines for the transfer of high-energy beams to the treatment rooms. The injector can be placed outside the ring for easier maintenance, or inside the ring to save space. It is made up of two or three sources (depending on the number of species one wants to use with short switching times), a Low-Energy Transfer Line (LEBT) to select and transport the ions produced by the sources, a linac, and the Medium-Energy Transfer Line (MEBT) to transport the preaccelerated beam to the synchrotron.

Different kinds of ion sources are used. Only a few facilities have a PIG source or an EBIS source. The PIG is based on Penning vacuum gauges: a flux of electrons between an anode and a cathode ionizes a gas, creating the beam. In the EBIS, ions are trapped inside a dense electron beam and are continuously bombarded by electrons and sequentially ionized. The most commonly used source is the ECR source. It is based on the excitement of electrons using radiofrequency fields at the electron cyclotron resonance frequency (10–18 GHz): the plasma electrons are confined by a magnetic trap (the so-called minimum B-structure) realized by the superposition of a hexapole and an axial structure. The advantage of an ECR source is that it produces high-intensity beams for a wide range of charge states.

The 14.5 GHz ECR sources [29] installed at CNAO can produce 250 $\mu$A of $C^{4+}$ beam with a normalized emittance of 0.56$\pi$ mm·mrad and 1000 $\mu$A of $H^{3+}$ with an emittance of 0.67$\pi$ mm·mrad with an energy of 8 keV/u. The RF can be finely adjusted in frequency, has a power of about 8 W for protons and 180 W for carbon ions, and is fed by 400 W TWTA power amplifiers.

The low-energy linac comprises an RFQ (Radio Frequency Quadrupole) and an IH (Interdigital H-type). The source current is d.c., but it enters in the RFQ pulsed thanks to an electrostatic chopper in the LEBT.

The CNAO RFQ [30] is of four-rod type with 70 kV electrode voltage, and delivers beams at 400 keV/u; the IH [30] is a 5.3 MV/m, 3.77 m long structure and accelerates beams up to 7 MeV/u; both tanks work at 217 MHz. The MEBT contains elements to fit the transverse and longitudinal twiss parameters to the ring acceptance, in addition to stripping foils that change the charge status from that produced by the sources. In the CNAO, stripping foils allow $C^{6+}$ and $H^+$ to be obtained; quadrupoles match the transverse dimensions, while a debuncher tank reduces the beam momentum spread.

In the required energy range, the proton magnetic rigidity varies in the range 1.16–2.31 T·m, and that for carbon ions ranges from 3.18 to 6.34 T·m. The use of normal conducting magnets giving at maximum 1.5 T implies that the ring length is about 60–80 m (the CNAO ring is 78 m). The acceleration is usually performed by a single cavity that must be a broadband resonator loaded with standard ferrites or with a ferrite-like amorphous alloy (CNAO uses VITROVAC, a Fe–Co alloy). Such alloys have several advantages, for example they reduce cavity dimensions and reduce (and in some cases eliminate) the current for the cavity polarization [31]. Extraction from the ring is the most important and challenging aspect influencing ring design. Clinical requirements on dose uniformity are ±2–3%; with active scanning, this requirement cannot be fulfilled with a single-turn extraction. A single-turn extraction means a beam shorter than 1 $\mu$s, requiring a passive system. As a consequence, a slow extraction of the order of 1 s is mandatory. The slow extraction mechanism is realized by making the particle betatron oscillations unstable: the amplitude of their motion grows steadily until the particle 'jumps' into the aperture of an electrostatic septum, allowing the extraction. The lattice layout of the ring must be set such that the machine tune at the end of the extraction is near to an unstable value: to extract the beam, a mechanism must force the beam into the unstable region. Essentially, three possible mechanisms are employed to make the beam pass from the stable to the unstable region. These are amplitude selection, amplitude–momentum selection, and RF Knock Out (RFKO). With amplitude selection, which is used in the oldest facilities, the quadrupole settings are changed before the extraction to vary the machine tune. In this case the beam that has a small momentum spread and a large betatron amplitude acquires the extraction tune progressively.

The beam size, position, and energy change during the extraction because only one amplitude is extracted at a time. In the amplitude–momentum selection process, the resonance region is fixed and the beam moves towards the resonance. Thus the momentum spread of the circulating beam is kept large, and the extracted beam has fixed position, size, and energy. At CNAO, the beam is driven into the resonance by a betatron core: it is a toroidal magnet that creates a fem that accelerates the beam towards the instability.

Finally, in the case of the RFKO method, the machine tune is fixed and the beam is excited by a transverse RF perturbation; size, position, and energy are stable. Furthermore, with this method it is easy to obtain a rapid switch off of the dose irradiation. Figure 10 illustrates graphically the three methods using the so-called 'Steinbach' diagram in which the resonance is represented in the phase-space betatron amplitude–momentum spread.

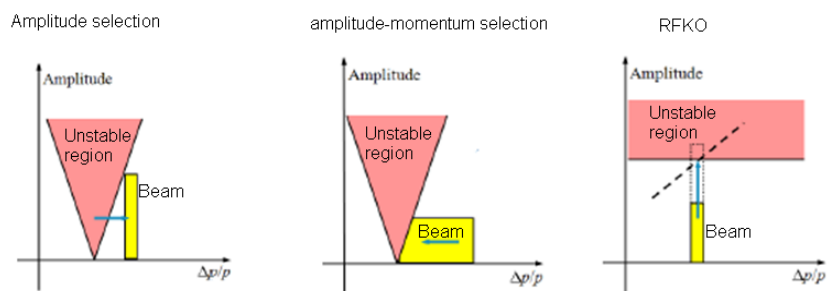

**Fig. 10:** Steinbach diagrams of the three methods that put the beam in resonance

At LLUMC, the unstable tune was chosen to be half-integer; the chosen resonance in all of the facilities is obtained by a tune of *N*/3 and a sextupolar field that feeds the instability (the so-called 'third integer resonance'). Another important aspect of the extraction is the intensity quality of the spill. Considering the 2% dose homogeneity and that the time to irradiate a voxel is about 5 ms, the beam has to be managed in a time structure of about 10 $\mu$s: this means that the spill intensity spectrum must be controlled up to 10 kHz. This control is not easy with amplitude selection because it requires a challenging control on the quadrupole ripple; on the contrary, using the amplitude–momentum selection and the RFKO technique, the spill structure can be well controlled. At CNAO the spill ripple is greatly reduced by the use of the empty bucket technique, i.e. simply exploiting the RF cavity used for the acceleration; some further improvements can be obtained by a rapid air core quadrupole in feedback on the spill intensity.

Finally, the extraction lines are also technologically challenging. First, the number of lines must be high, with rapid switching among the lines, to maximize the number of patients. Second, the beam quality needed at all energies (stable position, with a possibility of having round beams with varying dimensions and so on) puts constraints on magnetic lattices and requires precise specifications for the power supplies, the magnets, the control system, the beam diagnostics that control in real time the dose delivered to the patient (the so-called 'nozzle'), and the patient positioning. In particular, the extraction lines must be equipped with a system able to guarantee a rapid switch off of the extracted beam (of the order of 100 $\mu$s considering the requirement on dose uniformity). Indeed, a rapid switch off is not possible with a betatron core that is a highly inductive element and then slow; also, in the RFKO case, the time of a switch off is of the order of 1 ms. At CNAO this is obtained by four fast chopper magnets (100 $\mu$s) (see Fig. 11) installed along the extraction line that create a bump on the beam orbit: if the bump is not performed, the beam orbit ends on a dump.

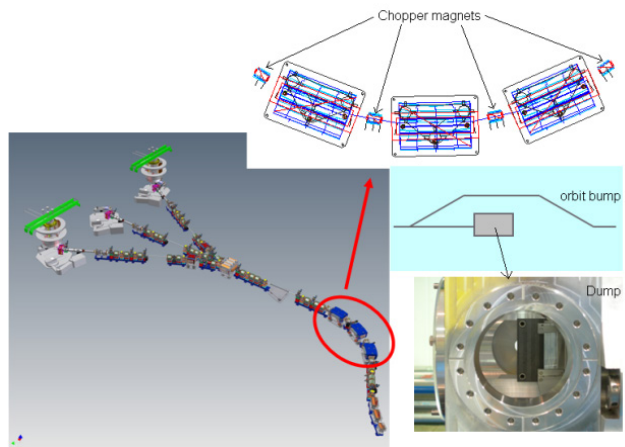

**Fig. 11:** CNAO safety system for a fast beam switch off

Irradiation from different directions is mandatory to improve the quality of the treatment. This is achieved either by displacing the patient, or by using several lines in the same room (e.g. horizontal and vertical) or installing rotating beam lines, the so-called gantries. Nowadays, gantries for protons are present in most facilities; on the contrary, a gantry for carbon ions must have a higher weight and size, and up to now only the Heidelberg facility is equipped with a carbon ion gantry (weighing 600 ton, with a diameter of 13 m compared with the standard dimensions for protons of 100 ton and 10 m) that is being commissioned [32].

### 5.4   Future developments and R&D

The hadrontherapy field is technologically challenging. Research centres contribute to the design and construction of facilities: CNAO is based on the PIMMS design [33] performed at CERN, and has been built with contributions from a network of international collaboration including INFN (Italy), CERN, GSI (Germany), LPSC (France), NIRS (Japan), and Italian universities (Milan, Pavia, Turin).

Beyond the R&D of technological aspects of the present hadrontherapy facilities, new ideas are being developed in the fields of tumour tracking, tumour imaging, and accelerator technology. We briefly mention some of the new technologies being addressed in the accelerator field.

The Fixed-Field Alternating Gradient (FFAG) design [34] foresees fixed-field combined-function bending magnets: a strong radial magnetic field gradient in the dipole component permits the beam to be kept in a narrow ring, as in a synchrotron, but with no need for magnet ramping. A d.c.

beam or at high rep rate with the possibility of fast energy changes is therefore feasible. Both scaling and non-scaling FFAGs have been proposed, the difference being how the field varies with radius. A non-scaling FFAG prototype, EMMA, is in operation at Daresbury [35], and is providing interesting results in terms of the available momentum range.

The Linac Booster (LIBO) [36] foresees a proton linac (1.5 m with 27 MV/m) booster from 30–250 MeV so it can be used in association with the standard cyclotron for radioisotopes; the application of this idea for carbon is under study.

The Dielectric Wall-induction Accelerator (DWA) idea is based on the use of new dielectrics able to sustain high voltage gradients of the order of 100 MV/m. A prototype is being developed at Livermore to be used for accelerating protons up to 150 MeV in a structure of less than 1 m [37].

The CYCLINAC is a novel protontherapy accelerator design based on the use of normal conducting RF linacs as boosters of a beam previously accelerated by a fixed-energy injector. This concept, i.e. the use of a cyclotron as an injector, is a long-term development concept of the TERA Foundation [38]. If the linac is small enough, it can be installed on a rotating support for a compact single-room facility, which is called TULIP [39], the TUrning LInac for Protontherapy. Another application is provided by the upgrade of existing 230–250 MeV proton cyclotrons to reach an energy of 350 MeV, enabling the implementation of proton radiography and high-energy proton therapy with a traversing beam, as proposed by PSI in the so-called IMPULSE project.

Finally, the concept of acceleration with high-power lasers is being developed in several laboratories [41, 42]; the details are beyond the scope of this work. The compactness of the laser accelerator makes the sizes of the proposed protontherapy systems similar to those of present conventional radiotherapies, including the flexibility in operating real-time PET verification. The proton beam would be produced in a target (solid or gas) that is illuminated by the laser and which can be easily rotated around the patient, as shown schematically in Fig. 12.

The future of this field in terms of R&D offers plenty of possibilities, which are as interesting and often closely related to the advances in accelerators dedicated to fundamental investigation.

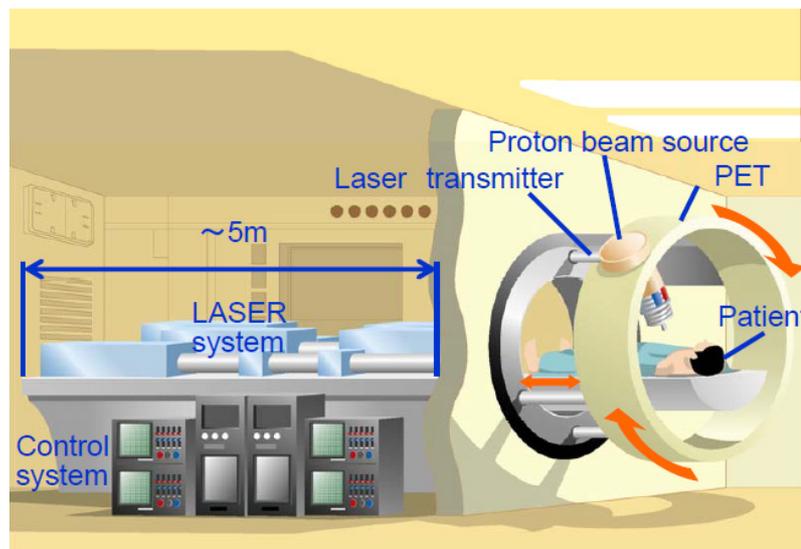

**Fig. 12:** Conceptual laser accelerated proton therapy instrument. Compact and flexible radiotherapy

## Acknowledgements

Special acknowledgements are due to Marco Pullia, Erminia Bressi, and Sandro Rossi from CNAO for precious references and for useful discussions.